\documentstyle[aps,epsfig]{revtex}
\begin{document}

\title{ On  the Experimental Identification of Spin-Parities and Single-Particle Configurations in $^{257}$No and Its $\alpha$-Decay Daughter $^{253}$Fm }

\author{P. Roy Chowdhury$^1$\thanks{E-mail:partha.roychowdhury@saha.ac.in} and D.N. Basu$^2$ }
\address{$^1$Saha Institute of Nuclear Physics, $^2$Variable  Energy  Cyclotron  Centre, 1/AF Bidhan Nagar, Kolkata 700 064, India}
\date{\today }
\maketitle
\begin{abstract}

      Recently measured lifetimes of the favored $\alpha$ decays from $^{257}$No have been calculated using the quantum mechanical tunneling within WKB approximation using microscopic nuclear potentials. Results obtained assuming previously assigned (ambiguous) parent spin of $\frac{7^+}{2}$ and newly assigned configuration $\frac{3^+}{2}$ [622] have been compared. Hindrance factors for the favored decays have also been compared with the calculated hindrances using higher angular momenta transfers. Although the calculations substantiate the findings, yet it makes the spin-parity assignment of $\frac{3^+}{2}$ for the ground state of $^{257}$No less definite. 

\end{abstract}

\pacs{ PACS numbers:23.60.+e, 21.30.Fe, 25.55.Ci }


      In a very recent experimental work \cite{As05}, lifetimes of the favored $\alpha$ decays from $^{257}$No have been measured and the spin-parities of the excited states in $^{253}$Fm fed by the $\alpha$ decays of $^{257}$No have been claimed to be identified on the basis of the measured internal conversion coefficients. In the present work, the decay $Q_{\alpha}$ values for the favored decays have been calculated from the measured $\alpha$ particle kinetic energy (K.E.) $E_{\alpha}$ using standard recoil correction and the electron shielding correction in a systematic manner as suggested by Perlman and Rasmussen \cite{Pe57}. The $Q_{\alpha}$ value for $\alpha$ decay from ground state (g.s.) of $^{257}$No to the g.s. of $^{253}$Fm has also been calculated using atomic mass excesses from the most recent experimental mass table \cite{Au03}. $Q_{\alpha}$ values, thus obtained, have then been used to calculate lifetimes of the favored as well as g.s. to g.s. $\alpha$ decays from $^{257}$No using the quantum mechanical tunneling within WKB approximation using microscopic nuclear potentials \cite{Ba03}. The nuclear interaction potentials required to describe these $\alpha$ decay processes have been calculated by double folding the density distribution functions of the $\alpha$ particle and the daughter nucleus with density dependent M3Y effective interaction. The microscopic $\alpha$-nucleus potential thus obtained, along with the Coulomb interaction potential and the minimum centrifugal barrier required for the spin-parity conservation, have been used for the lifetime calculations of these $\alpha$ disintegration processes. Spherical charge distributions have been used for calculating the Coulomb interaction potentials \cite{Ba03}. These calculations provide good estimates for the observed $\alpha$ decay lifetimes of nuclei including superheavies \cite{Ba04}.

      The decay $Q_{\alpha}$ values for the favored decays can be obtained from the measured $\alpha$ particle K.E. $E_{\alpha}$ using the following expression 

\begin{equation}
 Q_{\alpha} = (\frac{A_p}{A_p-4})E_{\alpha} + (65.3 Z_p^{7/5} - 80.0 Z_p^{2/5}) \times 10^{-6} ~\rm MeV
\label{seqn1}
\end{equation}
\noindent
where the first term is the standard recoil correction and the second term is an electron shielding correction in a systematic manner as suggested by Perlman and Rasmussen \cite{Pe57}. But for decays to the g.s. of daughter nucleus for which $\alpha$ particle K.E. has not been measured can be calculated from the g.s. masses using the following relationship 

\begin{equation}
 Q = M - ( M_\alpha + M_d)
\label{seqn8}
\end{equation}
\noindent
which being positive allows decay to the g.s., where $M$, $M_\alpha$ and $M_d$ are the atomic masses of the parent nucleus, the emitted $\alpha$-particle and the residual daughter nucleus, respectively,  expressed in the units of energy.

      The results of the present calculations have been presented in Table-I. The $Q_{\alpha}$ value of 8.466 MeV calculated for the transition to g.s. of $^{253}$Fm from the experimental mass excesses \cite{My96}, \cite{Au03} which should be more but turns out to be less than the the value of 8.496(7) MeV derived \cite{Pe57} from the measured $\alpha$ particle K.E. for transition to the excited state at 22.3 keV. This experimental result for $E_\alpha$ appears to be slightly erroneous. However, it is interesting to note that if only the recoil correction is considered then the result is 8.455(7) MeV which is almost within the acceptable limits. The theoretical results for the half-lives are often underestimated because the centrifugal barrier required for the spin-parity conservation could not be taken into account due to non availability of the spin-parities of the parent or the daughter/decay chain nuclei. The term $\hbar^2 l(l+1) / (2\mu r^2)$, where $\mu$ is the reduced mass and r is the distance between $\alpha$ and daughter nuclei, represents the additional centrifugal contribution to the barrier that acts to reduce the tunneling probability if the angular momentum carried by the $\alpha$-particle is non-zero. Hindrance factor which is defined as the ratio of the experimental mean life $\tau_{expt.}$ (or experimental half life) to the theoretical mean life $\tau_{th.}$ (or theoretical half life) is therefore larger than unity since the decay involving a change in angular momentum can be strongly hindered by the centrifugal barrier. The hindrance factor (HF) 

\begin{equation}
 HF = \tau_{expt.}/\tau_{th.},
\label{seqn3}
\end{equation}
\noindent
however, can also be different from unity because of other factors (such as nuclear deformations) not considered in theoretical calculations. But these effects are much smaller. 

      The total mean life $\tau$ (or half life $T_{1/2}$) can be obtained from the partial mean lives $\tau_1$, $\tau_1$, $\tau_2$, $\tau_3$, $\tau_4$ . . . . .  (or partial half lives) using 

\begin{equation}
 \frac{1}{\tau} = \frac{1}{\tau_1} + \frac{1}{\tau_2} + \frac{1}{\tau_3} + \frac{1}{\tau_4} . . . . . 
\label{seqn4}
\end{equation}
\noindent
which suggests that the total lifetime is always less than or equal to any of the partial lifetimes. 
 
\begin{table}
\caption{ Calculated half lives for $\alpha$ decays from the ground state of $^{257}$No to different excited states and to the ground state of  $^{253}$Fm. Calculated $Q_{\alpha}$ values have been derived from measured $\alpha$ particle K.E. $E_{\alpha}$ using eqn.(1) [2] whereas the $Q_{\alpha}$ value for $\alpha$ decay to the g.s. has been calculated from the atomic masses [3] using eqn.(2). The minimum angular momenta l$\hbar$ carried away by the $\alpha$ particles have been decided by the spin-parity coservation. Uncertanties in the calculated half lives arising from the experimental uncertainties in the $Q_{\alpha}$ values and uncertain spins have been provided within parentheses. }

\begin{tabular}{ccccccccc}
Parent &Ground state&Daughter &Excited state& Excited state&Measured $E_{\alpha}$&Derived $Q_{\alpha}$&$T^{th}_{1/2}$ & l      \\ 
 
nucleus&$J^{\pi}$&nucleus&Energy (MeV)&$J^{\pi}$&MeV&MeV&s & $\hbar$ \\ \hline

$^{257}$No  &($\frac{7^+}{2})$ \cite{Fi99} & $^{253}$Fm &g.s.&$\frac{1^+}{2}$&.....&8.46\cite{My96}&73.95&4     \\

$^{257}$No &($\frac{7^+}{2})$ \cite{Fi99}& $^{253}$Fm&g.s.&$\frac{1^+}{2}$&.....&8.466\cite{Au03}&69.78&4    \\  

$^{257}$No  &$\frac{3^+}{2}$ \cite{As05}& $^{253}$Fm&g.s.&$\frac{1^+}{2}$&.....&8.46\cite{My96}&17.95&2\\

$^{257}$No  &$\frac{3^+}{2}$ \cite{As05} & $^{253}$Fm&g.s.&$\frac{1^+}{2}$&.....&8.466\cite{Au03}&16.84&2\\  

$^{257}$No&$\frac{3^+}{2}$\cite{As05} &$^{253}$Fm&0.1241 &$\frac{3^+}{2}$&8.222(6)&8.394(6)&15.68(72)&0   \\ 

$^{257}$No&$(\frac{7^+}{2})$\cite{Fi99}&$^{253}$Fm&0.1241&$\frac{3^+}{2}$&8.222(6)&8.394(6)&28.94(133)&2   \\ 

$^{257}$No&$\frac{3^+}{2}$\cite{As05}&$^{253}$Fm&0.0223 &$(\frac{3^+}{2}$)&8.323(7)&8.496(7)&7.28(38)&0\\ 

$^{257}$No&$(\frac{7^+}{2})$ \cite{Fi99}&$^{253}$Fm&0.0223&$(\frac{3^+}{2}$)&8.323(7)&8.496(7)&13.48(70)&2 \\ 

$^{257}$No&($\frac{7^+}{2})$\cite{Fi99}&$^{253}$Fm&0.0223&$(\frac{1^+}{2}?$)&8.323(7)&8.496(7)&55.47(291)&4 \\ 

\end{tabular}

\end{table}

     As the spin-parity of the state at 22.3 keV of $^{253}$Fm is uncertain \cite{As05}, transitions other than l = 0 may be possible. But angular momentum l = 2 carried away by the $\alpha$ particle for transiton to the 124.1keV state is not possible. Therefore, transition to 22.3 keV state with half life higher than 7.28(38) s with HF of 2 or 8 may be possible but the half life for the transition to 124.1 keV state can not be more than 15.68(72) s since spin-parity of this state has been claimed to be definite \cite{As05}. Since eqn.(4) shows that the total half life must be less than the most favored decay lifetime, present calculations somewhat underestimates the measured half life of 24.5(5) s. But certainly the earlier less definite assignment of the spin-parity of $\frac{7^+}{2}$ \cite{Fi99} for the $^{257}$No suggests g.s. to g.s. $\alpha$ emissions less favored since then the spin-parity conservation forces 4 units of angular momentum to be carried away by the $\alpha$ particle for transition to the ground state, resulting a too high lifetime of about 70 s. For transitions to the states at 124.1 keV and 22.3 keV with spin-parity $\frac{3^+}{2}$, from a state with $\frac{7^+}{2}$, the spin-parity conservation forces a minimum of 2 units of angular momentum to be carried away by the $\alpha$ particle, resulting the half-lives of 28.94 s and 13.48 s respectively. Moreover, half life of the 13.48 s may also be different as the spin $(\frac{3^+}{2})$ of the 22.3 keV level is also somewhat uncertain. The ENSDF half life value of 25(2) s for the favored $\alpha$ decay to $^{253}$Fm going to a level at $\approx$ 100 keV is reasonably close to the value of 28.94(133) s calculated assuming $\frac{7^+}{2}$ for the spin-parity for the ground state of $^{257}$No. Half lives of 28.94 s, 55.47 s (?) and  73.95 s may result in a measured half life of about 15 s. Therefore, from the measured $\alpha$ decay half life of 24.5(5) s, it is not possible to rule out the possibility of the g.s. spin-parity of $\frac{7^+}{2}$ for the $^{257}$No.

\end{document}